\documentclass[prl,aps,twocolumn,preprintnumbers,showpacs,showkeys,superscriptaddress]{revtex4}

\usepackage{epsfig,amssymb,amsfonts,verbatim}

\def\bfl{\begin{flushleft}}
\def\efl{\end{flushleft}}
\def\bfr{\begin{flushright}}
\def\efr{\end{flushright}}
\def\bc{\begin{center}}
\def\ec{\end{center}}
\def\be{\begin{equation}}
\def\ee{\end{equation}}
\def\ba{\begin{eqnarray}}
\def\ea{\end{eqnarray}}
\def\baa#1{\begin{array}{#1}}
\def\eaa{\end{array}}
\def\bw{\begin{widetext}}
\def\ew{\end{widetext}}

\begin{document}

\title{Comments on "Thermal conductivity of nanotubes: Effects of chirality and isotope impurity" [cond-mat/0501194]}

\author{Andrew Das Arulsamy}
\affiliation{Condensed Matter Group, Division of Exotic Matter,
No. 22, Jalan Melur 14, Taman Melur, 68000 Ampang, Selangor DE,
Malaysia}

\date{\today}

\begin{abstract}

Based on Zhang and Li's molecular dynamics calculations, the heat
conductivity ($\kappa$) of metallic single-wall Carbon nanotubes
(SNTs) was incorrectly concluded as mainly influenced by the
radius. Apart from that, their isotope effect interpretation on
$\kappa(T)$ for the metallic SNTs using solely the phonon
scattering mechanism is also highlighted to be incorrect.

\end{abstract}


 \maketitle


Recently, G. Zhang and B. Li (ZL) have attempted to propose a
model that explains how thermal conductivity is related to
chirality and isotope concentration by using the non-equilibrium
molecular dynamics~\cite{zl1}. In their calculations, they have
utilized the Hamiltonian with the Tersoff empirical
potential~\cite{tersoff2}, which is given by, $H = \sum_i
\big[(p_i^2/2m_i) + (1/2)\sum_{j,j\neq i}
f_c(r_{ij})\big(V_R(r_{ij}) + b_{ij}V_A(r_{ij})\big)\big]$. $f_c$
is a function of the atomic distance, $V_{R,A}$ denotes the
repulsive and attractive potentials respectively. Whereas $b_{ij}$
describes the bonding nature of atoms $i$ and $j$, or implicitly
describes how the valence electrons of Carbons influences the
phonons vibrational properties. The heat gradient was achieved
with the Nos\'{e}-Hoover heat baths~\cite{nose3,hoover4}. In
addition, the thermal current, $J = -\kappa\triangledown T$ is
used by assuming a steady state heat flow. It has been shown that
the $T$-profile along the nanotube, d$T$/d$x$ is identical between
(9,0)-metallic and (10,0)-semiconducting nanotubes. The data
listed in Table 1 of Ref.~\cite{zl1} are plotted in
Fig.~\ref{fig1}. From the definition, a chiral molecule cannot be
superimposed on its own mirror image~\cite{march7}. Therefore,
($n$,0) and $(n,n)$ are achiral SNTs. SNTs are usually
characterized by the index $(n,m)$, which originates from the
roll-up or chiral vector, $C_h = n\textbf{a}_1 + m\textbf{a}_2$
where $\textbf{a}_1$ and $\textbf{a}_2$ are the 2D graphite-sheet
lattice vectors while $n$ and $m$ are integers~\cite{odom8}. Refer
to Figs.~\ref{fig2} and~\ref{fig3}, which clearly point out that
the chirality exists for the chiral angle, $\zeta$ in the range of
0$^o < \zeta < 30^o$. Note that the Figs.~\ref{fig2}
and~\ref{fig3} were carbon-copied from the Ref.~\cite{weisman9}
and Ref.~\cite{odom8} respectively. In a first order
approximation, SNTs can be made metals if $(n-m)/3$ equals an
integer, otherwise they are semiconductors~\cite{odom8}. In other
words, (9,0) (5,5) are metals, whereas (10,0) is a semiconductor.

ZL should realize that the Landauer theory of purely phononic
scenario discussed by Yamamoto {\it et al}.~\cite{land10} is
limited to the semiconducting SNTs. As such, the total heat
conductivity for metallic SNTs should be written as $\kappa =
\kappa_e + \kappa_{ph}$~\cite{land10}. Consequently,
$\kappa^{\zeta = 0^o}_{(9,0)}
> \kappa^{\zeta = 0^o}_{(10,0)}$ and $\kappa^{\zeta = 30^o}_{(5,5)} > \kappa^{\zeta = 0^o}_{(10,0)}$ from
Table 1 of Ref.~\cite{zl1} do not indicate that the changes in
$\kappa$ as mainly from $r$ since $\kappa^{\zeta = 0^o}_{(9,0)}$
and $\kappa^{\zeta = 30^o}_{(5,5)}$ are also influenced by
$\kappa_e$ significantly. It is unclear and disturbing as to why
ZL have sidestepped the contribution of $\kappa_e$ even in the
metallic (9,0) and (5,5) SNTs.

Subsequently, interpreting the results plotted in Fig. 2b of
Ref.~\cite{zl1} based on phonon scattering mechanism alone is
absurd because (5,5) is likely a metallic SNT. Therefore,
$\kappa_e(T)$ is definitely a non-negligible term~\cite{land10},
complying with the arguments mentioned above. In simple words,
isotope effect pops out in the total heat conductivity as given in
Eqs.~(\ref{eq:1}) and~(\ref{eq:2}) below

\begin{eqnarray}
&&\kappa^{^{12}C}(T) = \kappa^{^{12}C}_{ph}(T) + \kappa_e(T).
\label{eq:1}\\&& \kappa^{^{12}C+^{14}C}(T) =
\kappa^{^{12}C+^{14}C}_{ph}(T) + \kappa_e(T).\label{eq:2}
\end{eqnarray}

These equations simply suggest that $\kappa_e(T)$ remains the
same, regardless of isotope doping. As a consequence, ZL should
first identify and isolate $\kappa_e(T)$ from Fig. 2b of
Ref.~\cite{zl1} in order to justify the phonon scattering
mechanism for $\kappa^{^{12}C}_{ph}(T)$ and
$\kappa^{^{12}C+^{14}C}_{ph}(T)$. Alternatively, they should
either show that the (5,5) SNT is a relatively large-gap
semiconductor or their Hamiltonian does not contain electrons,
which is meaningless for condensed matter applications anyway.
Otherwise, their analyses presented in the cond-mat/0501194 are
literally incorrect.

The author is grateful to Arulsamy Innasimuthu, Sebastiammal
Innasimuthu, Arokia Das Anthony and Cecily Arokiam of CMG-A for
their hospitality. I also thank Chong Kok Boon for providing some
of the references.

\begin{figure}
\caption{Calculated heat conductivity ($\kappa$) versus SNT's
radius based on Zhang and Li~\cite{zl1}'s calculations.}
\label{fig1}
\end{figure}

\begin{figure}
\caption{Single-walled carbon nanotubes exist in a variety of
structures corresponding to the many ways a sheet of graphite can
be wrapped into a seamless tube~\cite{weisman9}.} \label{fig2}
\end{figure}

\begin{figure}
\caption{Schematic of a 2D graphene sheet illustrating lattice
vectors $\textbf{a}_1$ and $\textbf{a}_2$, and the roll-up vector
$C_h = n\textbf{a}_1 + m\textbf{a}_2$, as well as the translation
vector $\textbf{T}$ along the nanotube axis that defines the 1D
unit cell.~\cite{odom8}.} \label{fig3}
\end{figure}


\begin{references}

\bibitem{zl1} G. Zhang, B. Li, cond-mat/0501194.

\bibitem{tersoff2} J. Tersoff, Phys. Rev. B 39 (1989) 5566; Phys.
Rev. Lett. 61 (1988) 2879; Phys. Rev. B 37 (1988) 6991; Phys. Rev.
B 38 (1988) 9902.

\bibitem{nose3} S. Nos\'{e}, J. Chem. Phys. 81 (1984) 511.

\bibitem{hoover4} W. G. Hoover, Phys. Rev. A 31 (1985) 1695.

\bibitem{march7} J. March, Advanced Organic Chemistry, Wiley, 1992.

\bibitem{odom8} T. W. Odom, J.-L. Huang, P. Kim, C. M. Lieber, J. Phys. Chem. B 104 (2000)
2794.

\bibitem{weisman9} R. B. Weisman, The Industrial Physicist (2002).

\bibitem{land10} T. Yamamoto, S. Watanabe, K. Watanabe, Phys. Rev. Lett. 92 (2004) 75502.

\end{references}
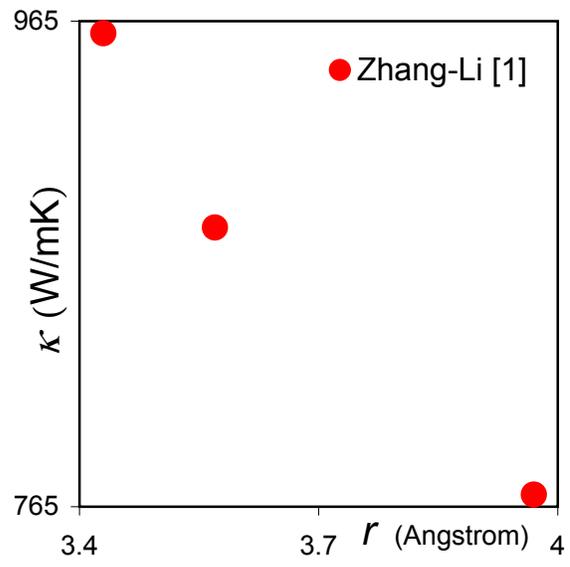
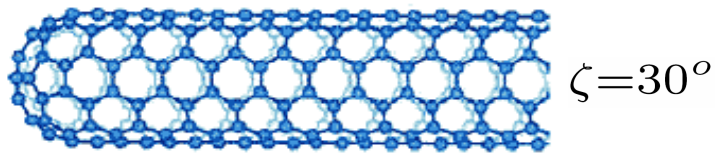
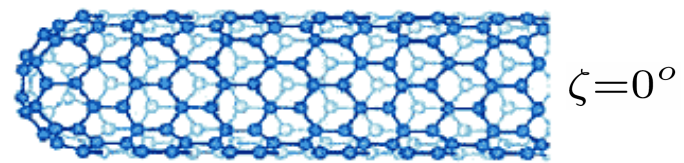
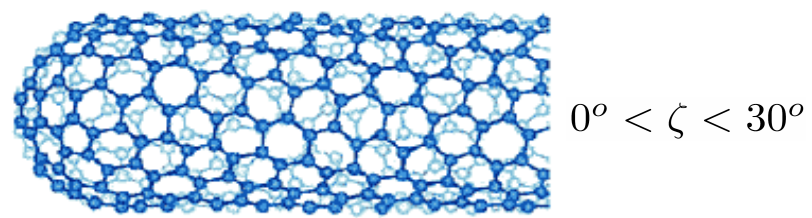
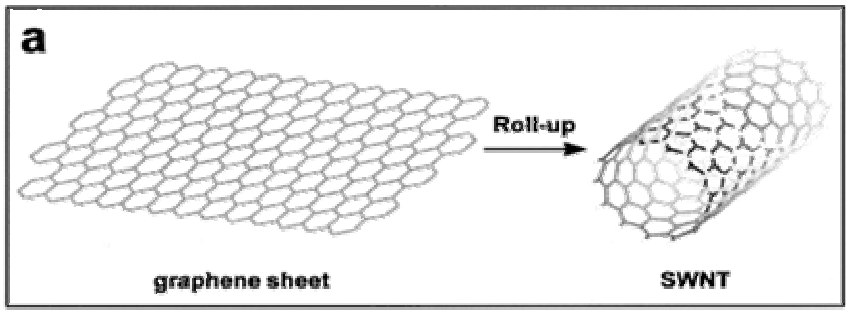
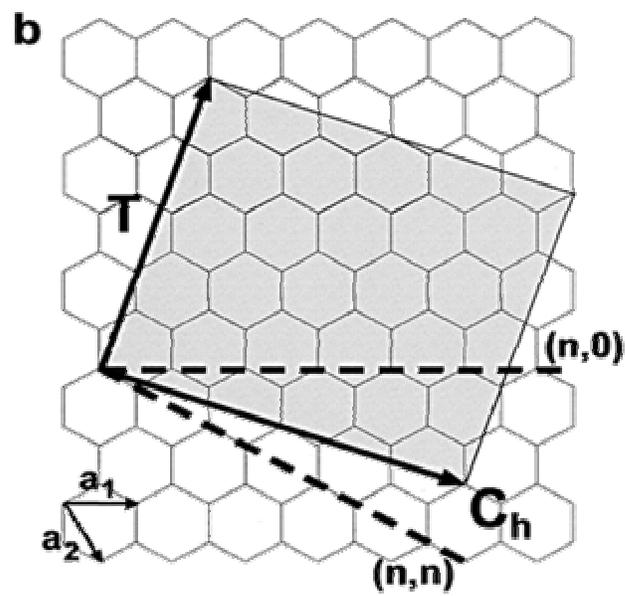
\end{document}